\documentclass[twocolumn,aps,prc,superscriptaddress,reprint]{revtex4-1}
\makeatletter

\newcommand{\Rmnum}[1]{\expandafter\@slowromancap\romannumeral #1@}
\makeatother
\usepackage{blindtext}
\usepackage{graphicx}
\usepackage{dcolumn}
\usepackage{bm}
\usepackage{booktabs}
\usepackage{mathrsfs}
\usepackage{ulem}
\usepackage{verbatim}
\usepackage{indentfirst}
\usepackage{float}
\usepackage{subfigure}
\usepackage[colorlinks,
           bookmarks=true,
           linkcolor=blue,
           urlcolor=blue, 
            anchorcolor=black,
            citecolor=blue
            ]{hyperref}
\usepackage{hypcap}
\usepackage{amsmath}
\newcommand{\sigs}{\sigma_{\rm soft}}

\usepackage{epsfig}

\begin{document}

\title{Update of a Multi-Phase Transport Model with Modern Parton
  Distribution Functions and Nuclear Shadowing}

\author{Chao Zhang}
\affiliation{Key Laboratory of Quark \& Lepton Physics (MOE) and Institute of Particle Physics,\\
Central China Normal University, Wuhan 430079, China}
\author{Liang Zheng}
\affiliation{China University of Geosciences, Wuhan 430074, China}
\affiliation{Key Laboratory of Quark \& Lepton Physics (MOE) and Institute of Particle Physics,\\
Central China Normal University, Wuhan 430079, China}
\author{Feng Liu}
\affiliation{Key Laboratory of Quark \& Lepton Physics (MOE) and Institute of Particle Physics,\\
Central China Normal University, Wuhan 430079, China}
\author{Shusu Shi}
\email{shiss@mail.ccnu.edu.cn}
\affiliation{Key Laboratory of Quark \& Lepton Physics (MOE) and Institute of Particle Physics,\\
Central China Normal University, Wuhan 430079, China}
\author{Zi-Wei Lin}
\email{linz@ecu.edu}
\affiliation{Key Laboratory of Quark \& Lepton Physics (MOE) and Institute of Particle Physics,\\
Central China Normal University, Wuhan 430079, China}
\affiliation{Department of Physics, East Carolina University, Greenville, NC 27858, USA}

\begin{abstract}
A multi-phase transport (AMPT) model has been successful in
explaining a wide range of observables in relativistic heavy
ion collisions. In this work, we implement 
a modern set of free proton parton distribution functions and 
an impact parameter-dependent nuclear shadowing in the AMPT model. 
After refitting the parameters of the two-component initial condition
model to the experimental data on $pp$ and $p \bar p$ total and inelastic 
cross sections from $\sqrt s \sim $ 4 GeV to 13 TeV, we study particle
productions in $pp$ and $AA$ collisions. 
We show that the updated AMPT
model with string melting can reasonably describe the
overall particle yields and transverse momentum spectra for both $pp$
and $AA$ collisions at RHIC and LHC energies 
after we introduce a nuclear scaling of the minijet transverse
momentum cutoff for $AA$ collisions at LHC energies that
is motivated by the color glass condensate. 
Since heavy flavor and high-$p_{\rm T}$ particles are produced by
perturbative-QCD processes and thus directly depend on parton
distribution functions of nuclei, the updated AMPT model is expected
to provide a more reliable description of these observables. 
\end{abstract}
\maketitle

\section{Introduction}

Experimental results from the Relativistic Heavy Ion Collider (RHIC) 
and the Large Hadron Collider (LHC) 
\cite{Arsene2005,Heinz:2013th,Busza:2018rrf} 
indicate that a hot and dense medium with partonic degrees
of freedom, namely the Quark-Gluon Plasma (QGP), is created in heavy
ion collisions at high energies. To study the properties of QGP, 
various theoretical methods and models are being developed including a
multi-phase transport (AMPT) model \cite{Lin:2004en}.  
The AMPT model aims to describe the whole phase space 
evolution of heavy-ion collisions as it contains four main
components: the fluctuating initial condition, partonic interactions,
hadronization, and hadronic interactions.  
The AMPT model has been widely used to simulate the evolution of the 
dense matter created in high energy heavy ion collisions. 
In particular, the string melting version of the AMPT
model~\cite{Lin:2001zk,Lin:2004en}, 
which converts the created matter in the overlap volume into parton
degrees of freedom, 
can well describe the anisotropic flows and particle correlations 
in collisions of small or large systems at both RHIC and LHC energies 
\cite{Lin:2004en,Lin:2001zk,Ma:2016fve,He:2017tla,Zhang:2018ucx}. 

However, the current public AMPT model (up to version
v1.26/v2.26~\cite{ampt}) uses the old Duke-Owens parton
distribution functions for the free proton and a schematic nuclear
shadowing parameterization from the HIJING~1.0 model 
~\cite{Wang:1991hta,Gyulassy:1994ew}. 
Therefore, it significantly underestimates
the gluon and quark distributions at small $x$. This would 
lead to significant uncertainties in its predictions on heavy flavor
and/or high-$p_{\rm T}$ observables, because those particles are initially
produced by perturbative-QCD processes and thus directly depend on the
parton distribution functions (PDFs) of nuclei. 
To improve the AMPT model for high energy nuclear collisions,
especially on heavy flavor and high-$p_{\rm T}$ observables, 
we incorporate in this study a modern set of free proton parton
distribution functions (the CTEQ6.1M set~\cite{Pumplin:2002vw}) 
and an impact parameter-dependent EPS09sNLO nuclear
shadowing~\cite{Eskola:2009uj} in an updated AMPT model.

The paper is organized as follows. 
After the introduction, 
we describe the initial condition of the AMPT model 
in section~\ref{sec:initial},
including the HIJING two-component model, 
the CTEQ6.1M parton distribution functions for the free proton,
the impact parameter-dependent EPS09s nuclear
shadowing functions, 
and our determination of the energy dependence of two key parameters
($p_0$ and $\sigs$) in the two-component model. 
We then investigate particle rapidity distributions and transverse
momentum spectra from the string melting version of the updated AMPT
model in Sec.~\ref{sec:results}, including our results for both $pp$ 
collisions and $AA$ collisions at  RHIC and LHC energies in
comparison with the experimental data. 
More discussions can be found in section~\ref{sec:discussions} 
including the effects of nuclear shadowing and the nuclear scaling of
the minijet transverse momentum cutoff $p_0$ on particle productions 
in $AA$ collisions. Finally, a summary is given in
section~\ref{sec:summary}. 

\section{The initial condition of the AMPT model}
\label{sec:initial} 

The string melting version of a multi-phase transport model
~\cite{Lin:2001zk,Lin:2004en} contains four main parts:  
the fluctuating initial conditions based on the HIJING two-component 
model~\cite{Wang:1991hta,Gyulassy:1994ew}, 
elastic parton scatterings modeled by the ZPC parton
cascade~\cite{Zhang:1997ej},
a spatial quark coalescence model to describe the hadronization 
of the parton matter~\cite{Lin:2001zk,He:2017tla}, 
and a hadron cascade based on the ART
model~\cite{Li:1995pra,Lin:2004en}. 
When we incorporate new parton distribution functions of nuclei in
the AMPT model, two key parameters in the HIJING
two-component model,  $p_0$ and $\sigs$, need to be retuned in
order to describe the cross sections of $pp$ and $p \bar p$ collisions.

\subsection{The HIJING two-component model}
\label{sec:hj}

The HIJING model~\cite{Wang:1991hta,Gyulassy:1994ew}, which combines
jet production that scales with the number of binary collisions with
string fragmentation, provides the initial condition of heavy ion
collisions in the AMPT model. 
In the HIJING model, the primary
interactions between the projectile and target are divided into soft
and hard components with a transverse momentum scale $p_0$.  
An interaction with a momentum transfer larger than $p_0$ is
considered to be a hard process and its production is calculated with 
perturbative QCD. 
On the other hand, 
the soft component with a momentum transfer below $p_{0}$  
is considered to be non-perturbative and characterized
by the cross section $\sigs$.

The inclusive jet differential cross section~\cite{Eichten:1984eu} in HIJING is determined by
\begin{align}\label{eq:definition2}
\frac{d\sigma_{\rm jet}}{dp_{\rm T}^{2}dy_{1}dy_{2}} = K\sum_{a,b}x_{1}f_{a}(x_1, Q^{2})x_{2}f_{b}(x_2, Q^{2})\frac{d\sigma^{ab}}{d\hat{t}},
\end{align}
where $p_{\rm T}$ is the transverse momentum transfer, $y_{1}$ and
$y_{2}$ are respectively the rapidity of the two produced partons,  
the $K$ factor aims to account for higher-order corrections, 
$x_{1}$ and $x_{2}$ are respectively the fraction of the momentum
carried by the two initial partons, 
$f_{a}(x_1, Q^{2})$ is the parton distribution function 
of parton type $a$ at the $x$-value of $x_1$ and factorization
scale $Q^{2}$, 
and $\sigma^{ab}$ is the
cross section between parton types $a$ and $b$.
The total inclusive jet cross section is then obtained by 
integrating the above with a transverse momentum cutoff $p_{0}$:
\begin{align}\label{eq:definition3}
\sigma_{\rm jet}(s)=\frac{1}{2}
  \int_{p_{0}^2}^{s/4}dp_{\rm T}^{2}dy_{1}dy_{2}\frac{d\sigma_{\rm jet}}{dp_{\rm T}^{2}dy_{1}dy_{2}}.
\end{align}

By introducing a soft interaction cross section $\sigs$, one can 
write an eikonal function~\cite{Gaisser:1984pg,Pancheri:1986qg} as 
\begin{align}\label{eq:definition4}
\chi(b,s)=\frac{1}{2}\sigs(s) T_{N}(b,s)+\frac{1}{2}\sigma_{\rm jet}(s) T_{N}(b,s), 
\end{align}
where $T_{N}(b,s)$ is the partonic overlap function between two
nucleons at impact parameter $b$ \cite{Wang:1991hta,Gyulassy:1994ew}. 
Then in the eikonal formalism, the total, elastic and inelastic cross
section of the nucleon-nucleon collisions can be written respectively as
\begin{eqnarray}
\sigma_{tot}&=&2\pi\int_{0}^{\infty}db^{2} \left [ 1-e^{-\chi(b,s)}
                \right ],\nonumber\\
\sigma_{el}&=&\pi\int_{0}^{\infty}db^{2} \left [ 1-e^{-\chi(b,s)} \right ]^2,\nonumber\\
\sigma_{in}&=&\pi\int_{0}^{\infty}db^{2} \left [1-e^{-2\chi(b,s)} \right ],
\end{eqnarray}
and they depend on both $p_0$ and $\sigs$.

\subsection{Parton distribution functions of the free proton}
\label{sec:pdf}

\begin{figure}[!htb]
 \begin{center}
   \includegraphics[scale=0.43]{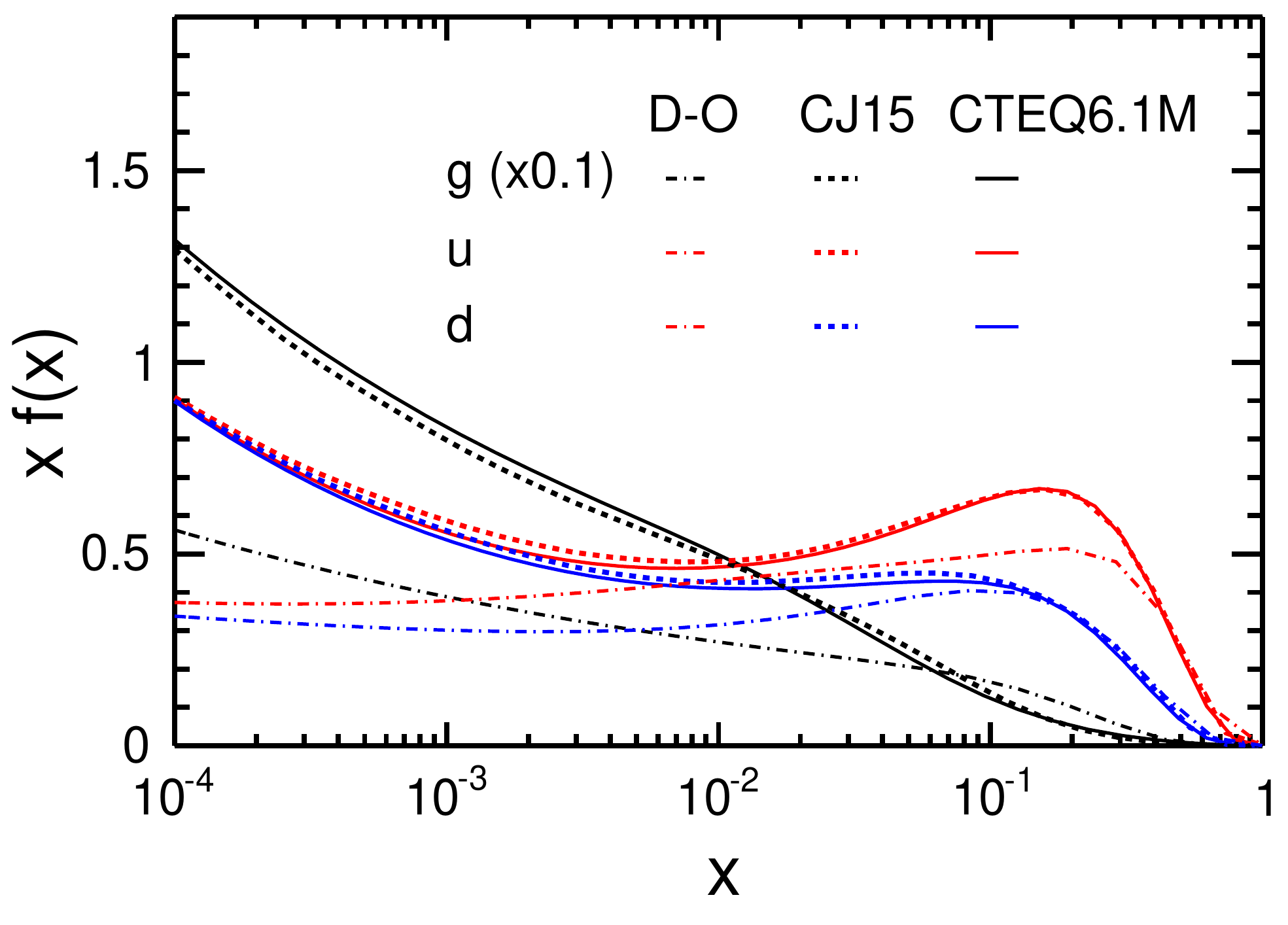}
   \caption{Parton density distributions of the free proton from the 
   CTEQ6.1M set (solid curves) in comparison 
   with the Duke-Owens (dot-dashed) and CJ15 (dashed) sets.
   }
   \label{fig:PDFcomparing}
  \end{center}
 \end{figure}

The HIJING~1.0 model in the current AMPT model uses the Duke-Owens 
parton distribution function set 1~\cite{Duke:1983gd} for the
free proton. 
However, it is well known that the Duke-Owens PDFs are outdated,
especially when the minijet productions reach the small-$x$ region 
of the parton distributions at high energies \cite{Lin:2011zzg}. 
So in this work we implement the modern 
CTEQ6.1M set ~\cite{Pumplin:2002vw}
for the parton PDFs of free proton (and free neutron). 
A similar update of parton PDFs has been done for the HIJING model, 
where the GRV PDFs \cite{Gluck:1994uf}
were used in the updated HIJING~2.0 model \cite{Deng:2010mv} to
replace the Duke-Owens PDFs. 

Figure~\ref{fig:PDFcomparing} compares 
the parton density distributions (PDFs multiplied by $x$)
from the Duke-Owens, CTEQ6.1M, and CJ15 sets for 
the gluon, u-quark and d-quark. 
Note that the gluon distributions have been scaled down by a factor of
ten. 
We see that all three distributions in the CTEQ6.1M parametrization
are quite different from the old Duke-Owens set and are much higher at
small $x$ values. 
In addition, differences between the CTEQ6.1M PDFs and the more recent
CJ15 PDFs \cite{Accardi:2016qay} are quite small. 

\subsection{Parton distribution functions in a nucleus}

Nuclear shadowing functions describe the modifications of parton
distribution functions in a nucleus relative to a simple superposition 
of parton distribution functions in the nucleon. 
Since we will be interested in describing nucleus-nucleus collisions
at various impact parameters, we implement the impact
parameter-dependent EPS09sNLO nuclear shadowing
functions~\cite{Eskola:2008ca}. 
They describe the spatial dependence of nuclear PDFs (nPDFs) and are
based on data from deep inelastic lepton-nucleus scatterings, 
Drell-Yan dilepton productions, and specifically  
pion productions measured at RHIC~\cite{Eskola:2008ca} which improve 
the determination of the gluon densities.
Note that the EPS09sNLO set was calculated with the CTEQ6M set as the
free proton PDFs, which is almost equivalent in every respect to the
CTEQ6.1M set \cite{Stump:2003yu}. 

For an average bound proton in a nucleus, the distribution function of
parton flavor $i$ can be written as
\begin{align}\label{eq:definition9}
f_{i}^{p/A}(x,Q^{2}) \equiv  R_{i}^{A}(x,Q^{2})f_{i}^p(x,Q^{2}),
\end{align}
where $f_{i}^p(x,Q^{2})$ is the corresponding PDF in the free proton.  
Here $R_{i}^{A}(x,Q^{2})$ represents the spatially-averaged nuclear
modification or shadowing function, which mainly contains three
effects depending on the $x$ range: the shadowing effect,
anti-shadowing effect, and the EMC effect.
It is an integral of the spatially-dependent nuclear shadowing
function as given by
\begin{align}\label{eq:definition10}
R_{i}^{A}(x,Q^{2}) \equiv \frac{1}{A}\int d^2{\bf s}~T_{A}({\bf s})~r_{i}^{A}(x,Q^{2},{\bf s}). 
\end{align}
In the above, $T_{A}({\bf s})$ is the nuclear thickness function at 
transverse position ${\bf s}$, 
and $r_{i}^{A}(x,Q^{2},{\bf s})$ is the spatially-dependent 
nuclear shadowing.

\begin{figure}[!htb]
 \begin{center}
\includegraphics[scale=0.43]{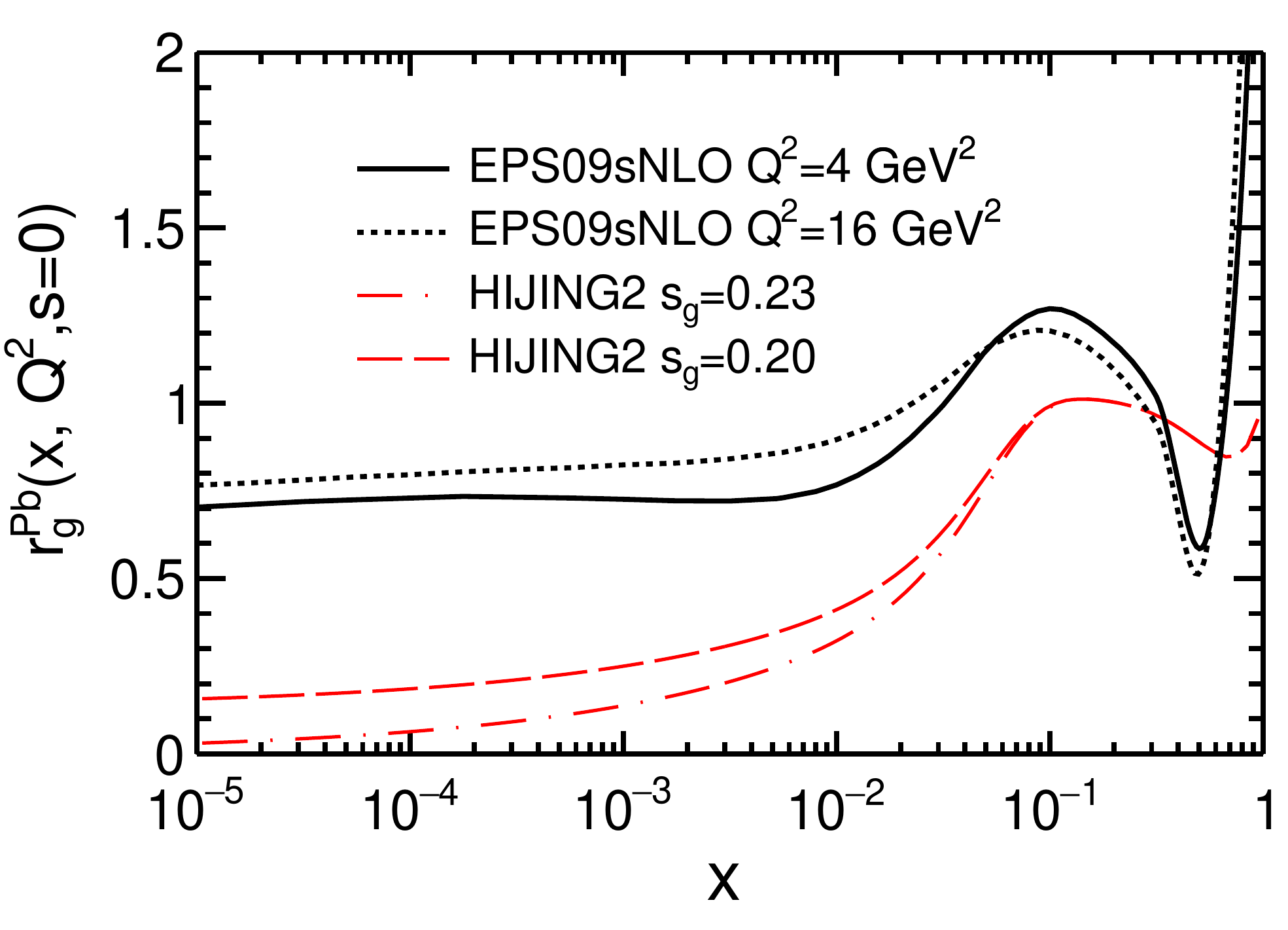}
\caption{Comparison of the nuclear shadowing functions of gluons 
at the center of a lead nucleus from the EPS09s NLO set at two different
$Q^2$ values and from the HIJING~2 parameterization with two different 
values for the $s_g$ parameter.
}
   \label{fig:shadowing}
  \end{center}
\end{figure}

Figure~\ref{fig:shadowing} shows the gluon shadowing functions at the
center of a lead nucleus from the EPS09s NLO set at two different
$Q^2$ values and from the HIJING~2.0 model at two different $s_g$
values suggested for LHC energies~\cite{Deng:2010mv}. 
We see that the EPS09sNLO gluon shadowing at small $x$ is much
weaker than the HIJING shadowing.
Note that the current AMPT model uses the HIJING~1.0 nuclear shadowing 
parametrization, which is spatially dependent but independent of $Q^2$
or the parton flavor~\cite{Wang:1991hta,Lin:2004en} and 
similar to the HIJING~2.0 nuclear shadowing.

\subsection{Fitting the two-component model to $pp$ and $p \bar p$ 
  cross section data} 
\label{sec:key}

The two parameters, $p_{0}$ and $\sigs$, in the HIJING~1.0 model 
directly affect the total and inelastic cross sections 
of $pp$ and $p \bar p$ collisions, as shown in Sec.~\ref{sec:hj}. 
In the current AMPT model that uses the Duke-Owens PDFs,
constant values of $p_{0}$ = 2.0 GeV$/c$ and $\sigs$ = 57 mb (at
high energies \cite{Wang:1990qp}) are found to be able to describe the
experimental cross sections of $pp$ and $p \bar p$
collisions~\cite{Wang:1991hta,Gyulassy:1994ew}. 
This is no longer the case after we use the CTEQ PDFs here, or when
the GRV PDFs were used for the HIJING~2.0 model~\cite{Deng:2010mv}. 
Instead, energy-dependent $p_{0}(s)$ and $\sigs(s)$
values are needed. 

\begin{figure}[!htb]
 \begin{center}
   \includegraphics[scale=0.43]{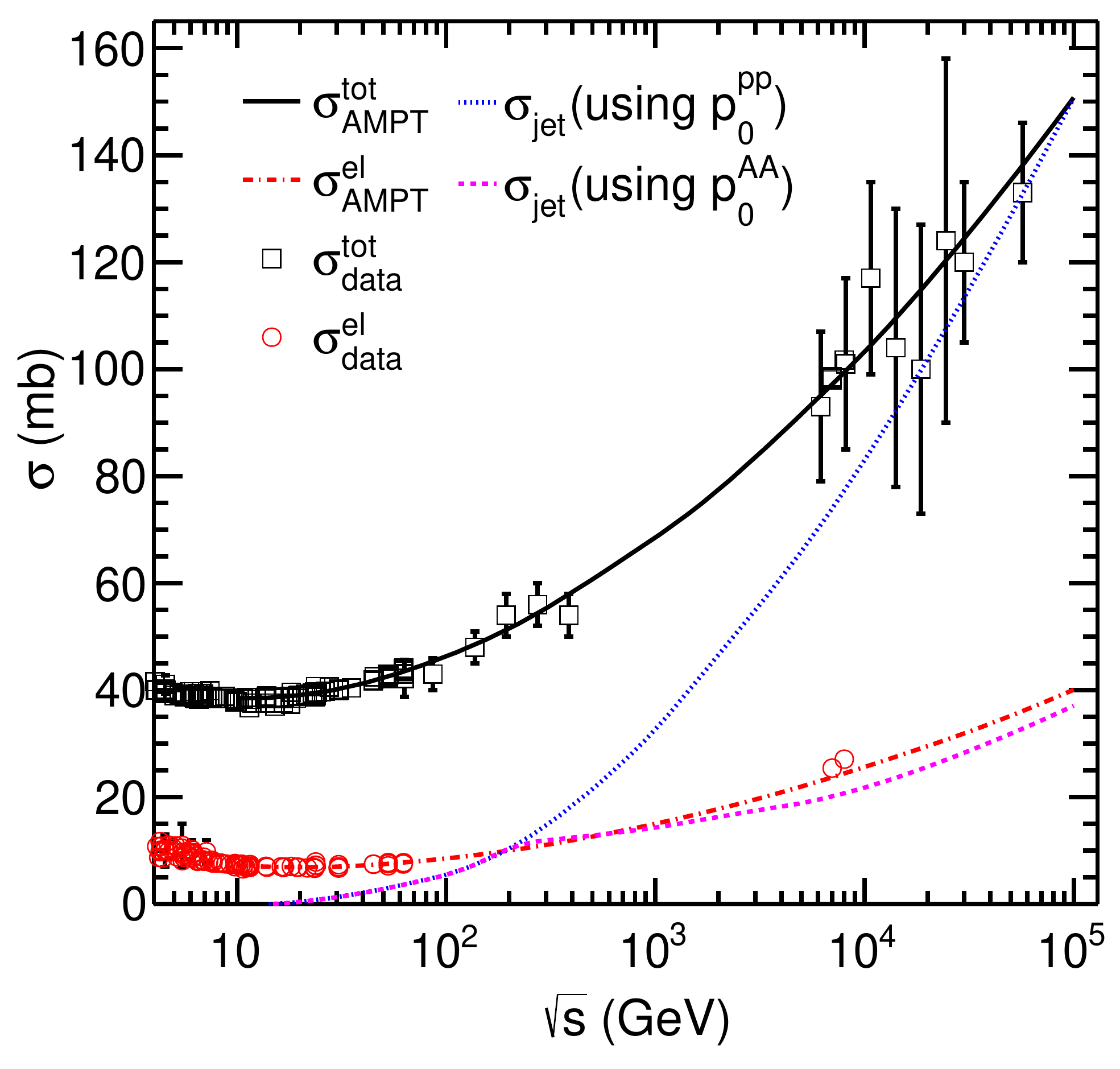}
   \caption{Total and elastic cross sections versus the colliding energy
     of $pp$ collisions from the experimental data
     (symbols) in comparison with the AMPT results (solid and
     dot-dashed curves); 
jet cross section per $pp$ collision, $\sigma_{\rm jet}$, is also shown for
$pp$ (dotted) and $AA$ (dashed) collisions.}
   \label{fig:parameterfit}
  \end{center}
 \end{figure}

Again we use the experimental total and inelastic cross sections 
of $pp$ and $p \bar p$ collisions within the energy range 
$4 < \sqrt s <  10^5$ GeV, as shown in Fig.~\ref{fig:parameterfit}, to
determine these two parameters at a given energy. 
To fit the experimental cross sections, 
we minimize the sum of squared relative difference 
between the model results and the cross section data points. 
We then determine the following fit functions of $p_{0}(s)$ and $\sigs(s)$:
\begin{eqnarray}
\label{eq:p0sigmas}
p_{0}^{pp}(s)=&-&1.71 +  1.63~ln(\sqrt{s})-0.256~ln^{2}(\sqrt{s})
\nonumber \\ 
 &+&0.0167~ln^{3}(\sqrt{s}),\\
  \sigs(s)=&&45.1 + 0.718~ln(\sqrt{s})+0.144~ln^{2}(\sqrt{s}) \nonumber \\ 
&+&0.0185~ln^{3}(\sqrt{s}).
\end{eqnarray}
In the above, $p_{0}^{pp}$ and $\sigs$ are in the unit of GeV$/c$ and
mb, respectively;  while the 
center-of-mass colliding energy $\sqrt{s}$ is in the unit of GeV.
Note that we have denoted the above minijet transverse momentum 
cutoff as $p_{0}^{pp}$ because it represents the $p_0$ fit
function for $pp$ collisions, while we shall see in Sec.~\ref{sec:NN} 
that $p_0$ needs to be $A$-dependent in order 
to reproduce the particle yields in $AA$ collisions at very high 
energies such as LHC energies. 
Also, $p_{0}$ values are only relevant when the center-of-mass 
energy per nucleon-pair is higher than 10~GeV, because the jet
production in the HIJING model is switched off at $\sqrt{s}<10$~GeV.

\begin{figure}[!htb]
 \begin{center}
   \includegraphics[scale=0.43]{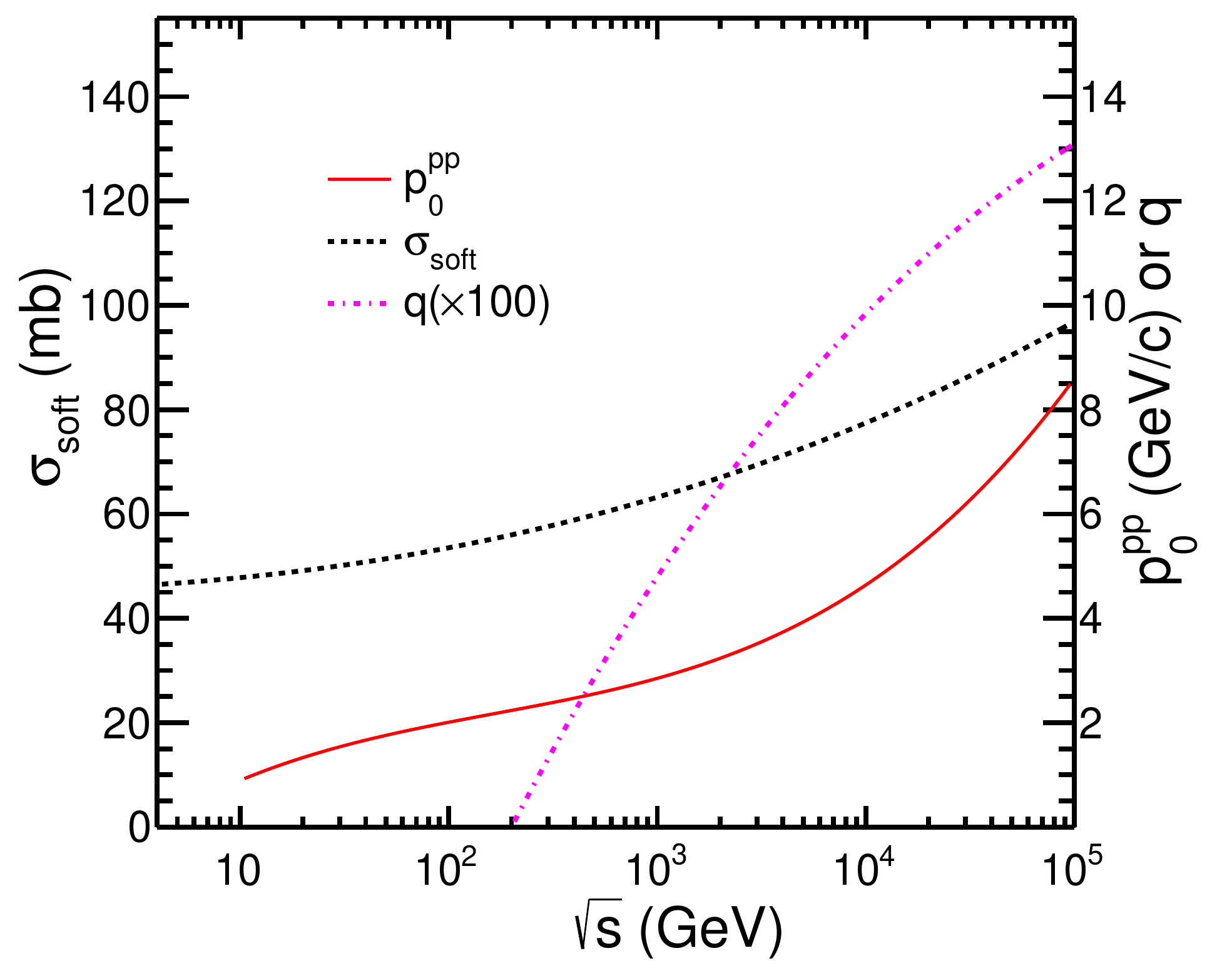}
   \caption{Fitted $p_{0}^{pp}$ function for $pp$ collisions (solid
     curve) and fitted $\sigs$ function (dashed) versus the colliding 
     energy; note that $p_{0}$ is only relevant at $\sqrt{s_{NN}} >
     10$~GeV. Dot-dashed curve represents the fitted q function
     (scaled up by a factor of 100) in the nuclear scaling of $p_0$ of
     Eq.~(\ref{eq:p0aa}).
   }
   \label{fig:p0_sigma}
  \end{center}
 \end{figure}

Figure~\ref{fig:p0_sigma} shows these two fit functions versus the
colliding energy. 
We see that both show a strong energy dependence, especially the
minijet cutoff scale $p_{0}$. 
Because the CTEQ parametrization has much higher gluon densities 
at small $x$ than the Duke-Owens PDFs, 
it has a larger jet cross section at high colliding energies,  
therefore a higher  $p_{0}$ value than the previous value of 2 GeV$/c$
is needed in order to reproduce the total and elastic cross section
data at high energies. 
As shown in Fig.~\ref{fig:parameterfit}, 
the above fit functions of $p_{0}(s)$ and $\sigs(s)$ allow the updated 
AMPT model to describe the experimental data on the total and elastic
cross sections of $pp$ collisions within a wide energy range 
$4< \sqrt s < 10^5$ GeV.

\section{Results on particle productions}
\label{sec:results}

We now study particle productions in $pp$ and $AA$ collisions with
the string melting version of the updated AMPT model and compare with
the experimental data. 
In the string melting AMPT model~\cite{Lin:2001zk,Lin:2004en},
the initial partons are produced through the intermediate step of 
Lund string fragmentation, where hadrons and resonances from the
fragmentation process are decomposed into (anti)quarks according to
the quark model. 
Therefore the initial phase-space distribution of the produced partons
depends on the string fragmentation parameters, particularly the $a$
and $b$ parameters in the  Lund symmetric fragmentation function: 
\begin{align}\label{eq:definition1}
f(z)\propto z^{-1}(1-z)^{a}~exp(-bm^{2}_{\rm T}/z).
\end{align} 
In the above, $z$ is the light-cone momentum fraction of the 
produced hadron with respect to the fragmenting string, and 
$m_{\rm T}$ is the hadron transverse mass. 
As a result, the final spectrum of produced particles in the AMPT
model depends on the Lund $a$ and $b$ parameters
~\cite{Lin:2004en,Lin:2014tya}. 
In particular, a smaller Lund $b$ value leads to a harder $p_{\rm T}$ 
spectrum~\cite{Lin:2014tya}. 
Note that the updated AMPT model used for this study also includes 
the new quark coalescence~\cite{He:2017tla}, which respects the
net-baryon conservation in each event but does not force the 
numbers of mesons, baryons, and antibaryons in an event to be
separately conserved through the quark coalescence process.

In this section, we first investigate particle productions 
in $pp$ collisions at RHIC and LHC energies 
to determine the values of the Lund $a$ and $b$ parameters.
We then apply the same Lund $a$ and $b$ values as well as the same
minijet cutoff value $p_0$ to $AA$ collisions,  
and we shall see that they fail to describe the experimental data of
central $AA$ collisions. 
We then keep the same Lund $a$ value but determine the Lund $b$ value
and the $A$-scaled $p_0$ value that are needed for the string melting AMPT
model to reproduce the overall particle productions in central $AA$
collisions at RHIC and LHC energies.

\subsection{Particle productions in $pp$ collisions}
\label{sec:pp}

\begin{figure}[!htb]
 \begin{center}
   \includegraphics[scale=0.43]{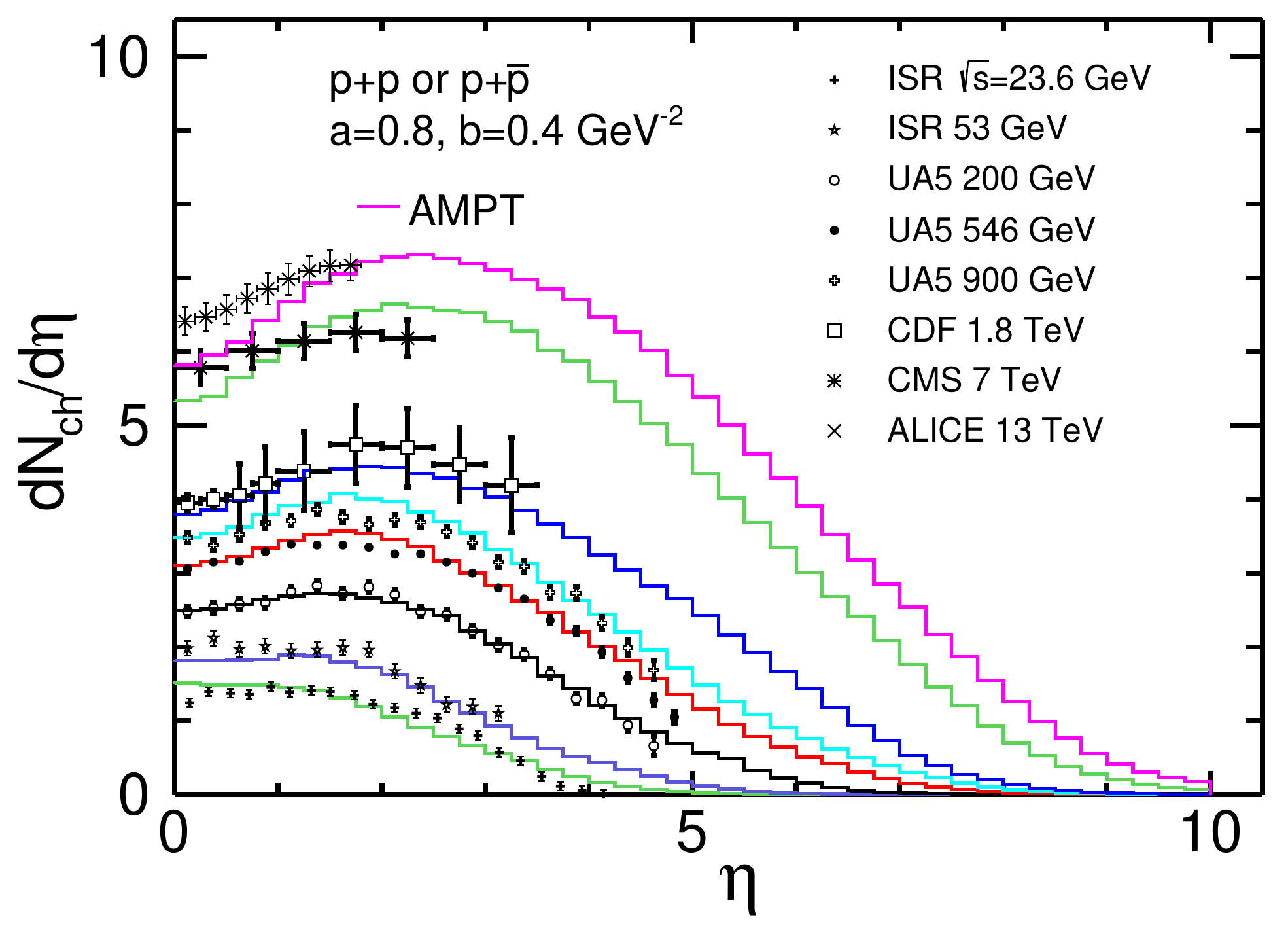}
   \caption{Pseudo-rapidity distributions of charged particles in
  NSD $pp$ collisions at $\sqrt{s}=7$ TeV~\cite{Khachatryan:2010us},
  NSD $p\bar p$ collisions at  $200, 546, 900$ ~\cite{Alner:1986xu}
  and $1800$ GeV~\cite{Abe:1989td}, and inelastic $pp$ collisions at
  23.6 GeV, 53 GeV~\cite{Thome:1977ky} and 13
  TeV~\cite{Khachatryan:2015jna} from AMPT (curves)  in comparison
  with the experimental data.
   }
   \label{fig:eta_pp}
  \end{center}
 \end{figure}

With the $p_0^{pp}(s)$ minijet cutoff function, 
using constant Lund fragmentation parameters of $a=0.8$ and 
$b=0.4$~GeV$^{-2}$ allows the string melting AMPT model to 
reasonably describe the $pp$ and $p\bar p$ data in 
both the $dN_{ch}/d\eta$ distributions and the $p_{\rm T}$ spectra. 
Figure~\ref{fig:eta_pp} shows charged particle 
pseudo-rapidity distributions from the updated AMPT model in
comparison with the experimental data 
of $pp$ and $p\bar p$ collisions from $\sim$20~GeV to 13~TeV. 
Note that we use the same procedure to select events for the AMPT
analysis  as that used for the experimental data.
For example, NSD events in the UA5 data refer to events that contains at
least one hit simultaneously on both sides of the chambers covering 
$2<|\eta|<5.6$, 
while for the CDF and CMS data they refer to the ranges of 
$3.2<|\eta|<5.9$ and $2.9<|\eta|<5.2$, respectively. 

\begin{figure}[!htb]
 \begin{center}
   \includegraphics[scale=0.43]{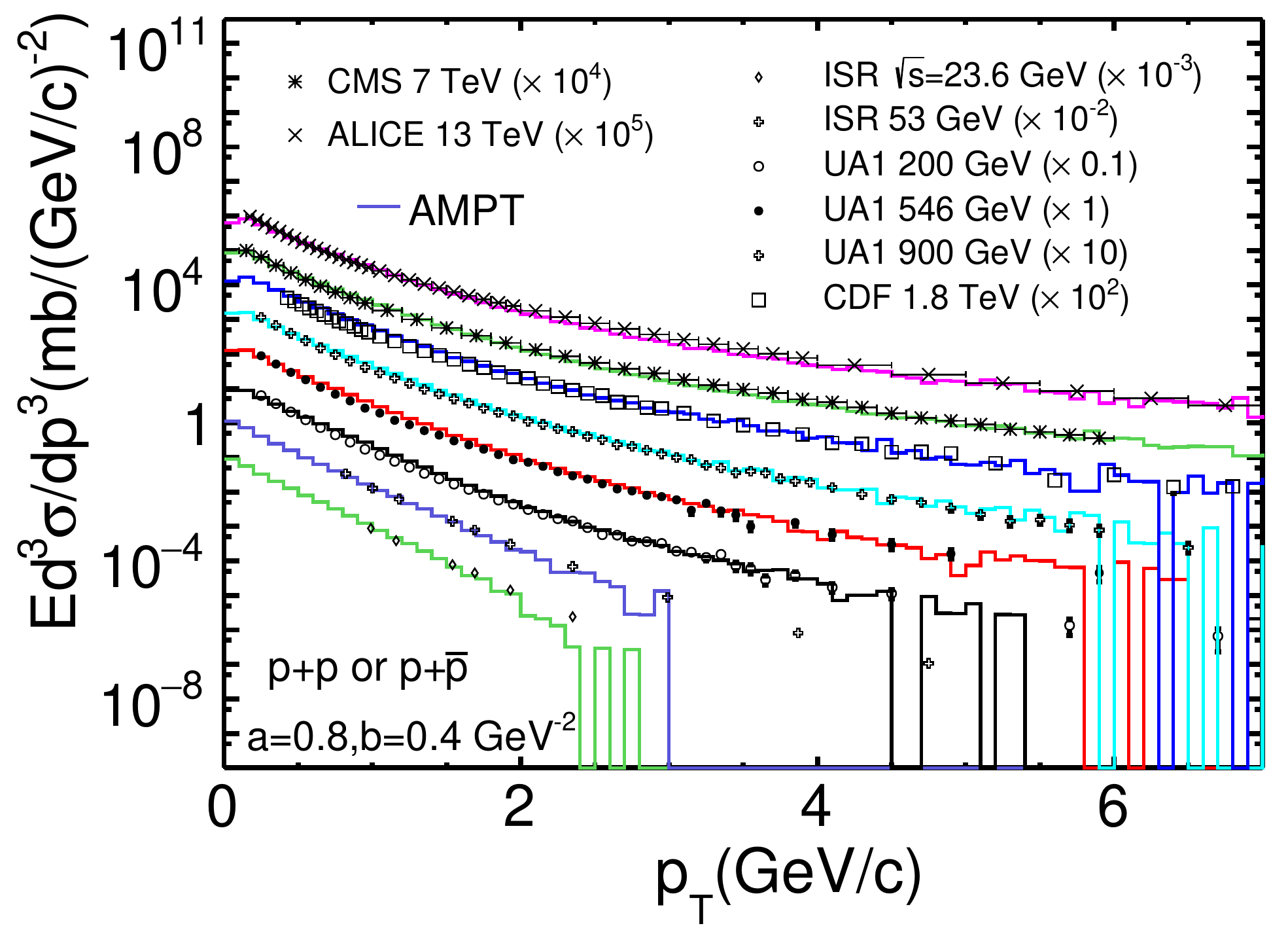}
   \caption{Invariant cross sections of charged particles 
versus $p_{\rm T}$ in $pp$ collisions at $\sqrt{s}$ = 23.6 and 53
     GeV~\cite{Thome:1977ky},  $p\bar{p}$ collisions at 200, 546,
     900~\cite{Albajar:1989an} and 1800 GeV~\cite{Abe:1988yu}, 
     and $pp$ collisions at 7~\cite{Abelev:2013ala} and 13
     TeV~\cite{Adam:2015pza}   from the AMPT model in comparison with
     the    experimental data. 
}
   \label{fig:pt_pp}
  \end{center}
 \end{figure}

Figure~\ref{fig:pt_pp} shows the transverse momentum spectra of
charged particles in $pp$ and  $p\bar p$ collisions from the string
melting AMPT model at different colliding energies in comparison with
data. Note that for $\sqrt{s}$ = 7 and 13~TeV, we have converted the
data on $E d^3N/dp^3$ to $E d^3\sigma/dp^3$.  
We have used the same $\eta$ range in calculating the AMPT results as
that in the experimental data:  
$|\eta|<0.35$ for $\sqrt{s} =$ 23.6 and 53~GeV, $|\eta|<2.5$ for
200, 546, and 900~GeV, $|\eta|<1$ for 1.8~TeV, $|\eta|<0.8$ for
7~TeV, and $|\eta|<2.4$ for 13~TeV.

\begin{figure}[!htb]
 \begin{center}
   \includegraphics[scale=0.43]{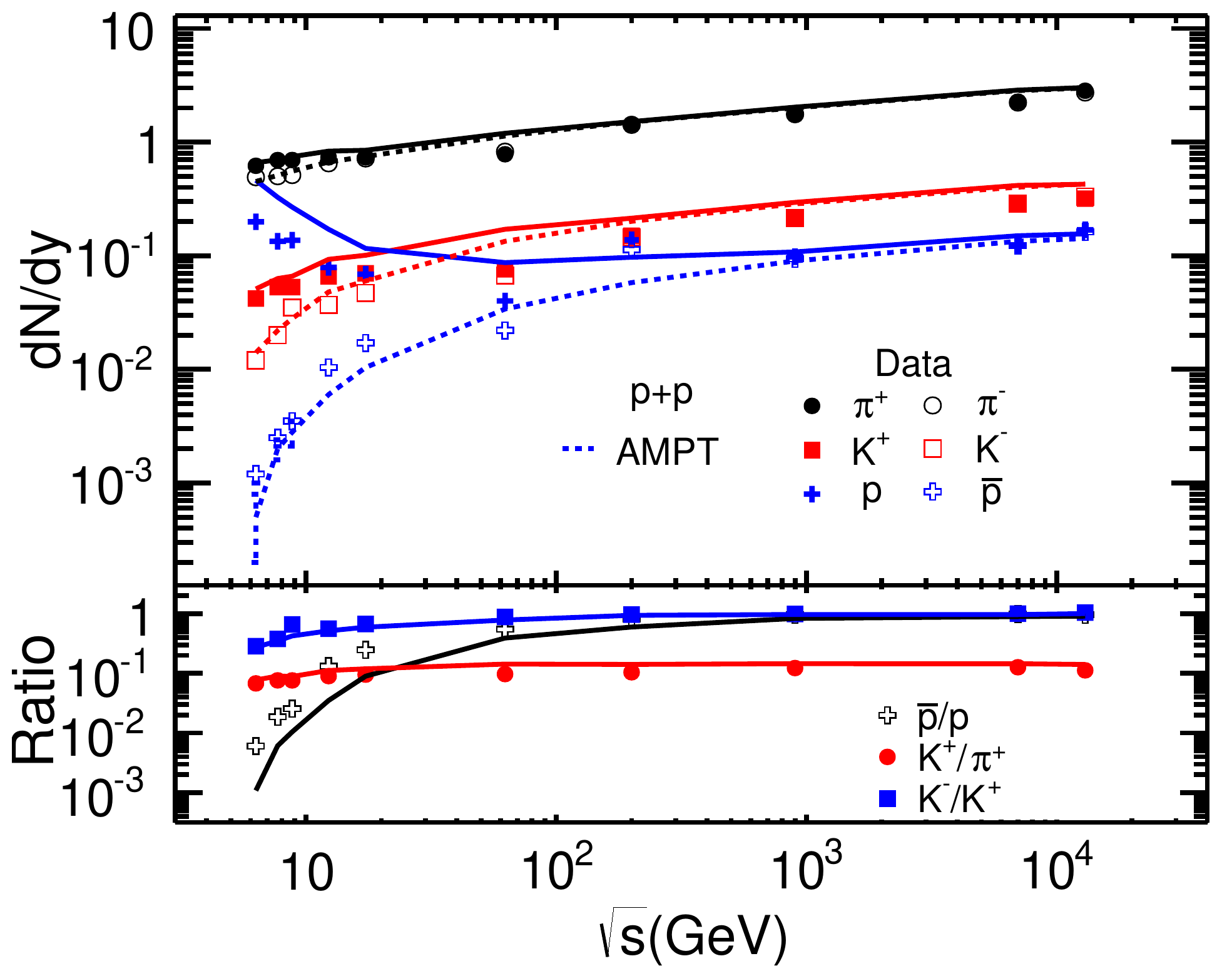}
   \caption{Identified particle $dN/dy$ 
and particle ratios at mid-rapidity in $pp$ collisions 
versus the colliding energy; 
curves represent the AMPT results with $a=0.8$ and $b=0.4$~GeV$^{-2}$,
while symbols represent the experimental data from 
NA61SHINE~\cite{Aduszkiewicz:2017sei}, PHENIX~\cite{Adare:2011vy},
STAR~\cite{Adams:2003qm} and
ALICE~\cite{Aamodt:2011zj,Adam:2015qaa,Sirunyan:2017nbr}.
      }
   \label{fig:pikp_pp}
  \end{center}
 \end{figure}

For identified particles, we compare in the upper panel of
Fig.~\ref{fig:pikp_pp} the string melting AMPT results 
on $dN/dy$ at mid-rapidity for pions, kaons, protons and
anti-protons in $pp$ collisions as functions of the colliding energy
from 6~GeV to 13~TeV. 
The experimental data are shown by symbols for comparison. 
We see that the string melting AMPT model can reasonably describe
the energy dependence of most of these hadrons, including the fast
increase of the antiproton 
yields with the colliding energy and the non-monotonous energy
dependence of the proton $dN/dy$. 

We also see from Fig.~\ref{fig:pikp_pp} that 
charged pion and kaon productions from the AMPT
model show good consistency with 
the $pp$ experimental data at different colliding energies, 
including the $K^+/\pi^+$ and $K^-/K^+$ ratios as functions of the
colliding energy as shown in the lower panel of Fig.~\ref{fig:pikp_pp}. 
However, the AMPT model here underestimates the antiproton yield 
and overestimates the proton yield at lower colliding energies. 
As a result, the $\bar p/p$ ratios from the AMPT model are lower than 
the data at the lower RHIC energies.
Note that in Fig.~\ref{fig:pikp_pp} the PHENIX proton and antiproton 
data~\cite{Adare:2011vy} shown at 62.4~GeV are corrected for
feed-down effects, but the STAR proton and antiproton data
~\cite{Adams:2003qm} shown at 200~GeV are not.

\subsection{Particle productions in AA collisions}
\label{sec:NN}

\begin{figure*}[!htb]
 \begin{center}
   \includegraphics[scale=0.9]{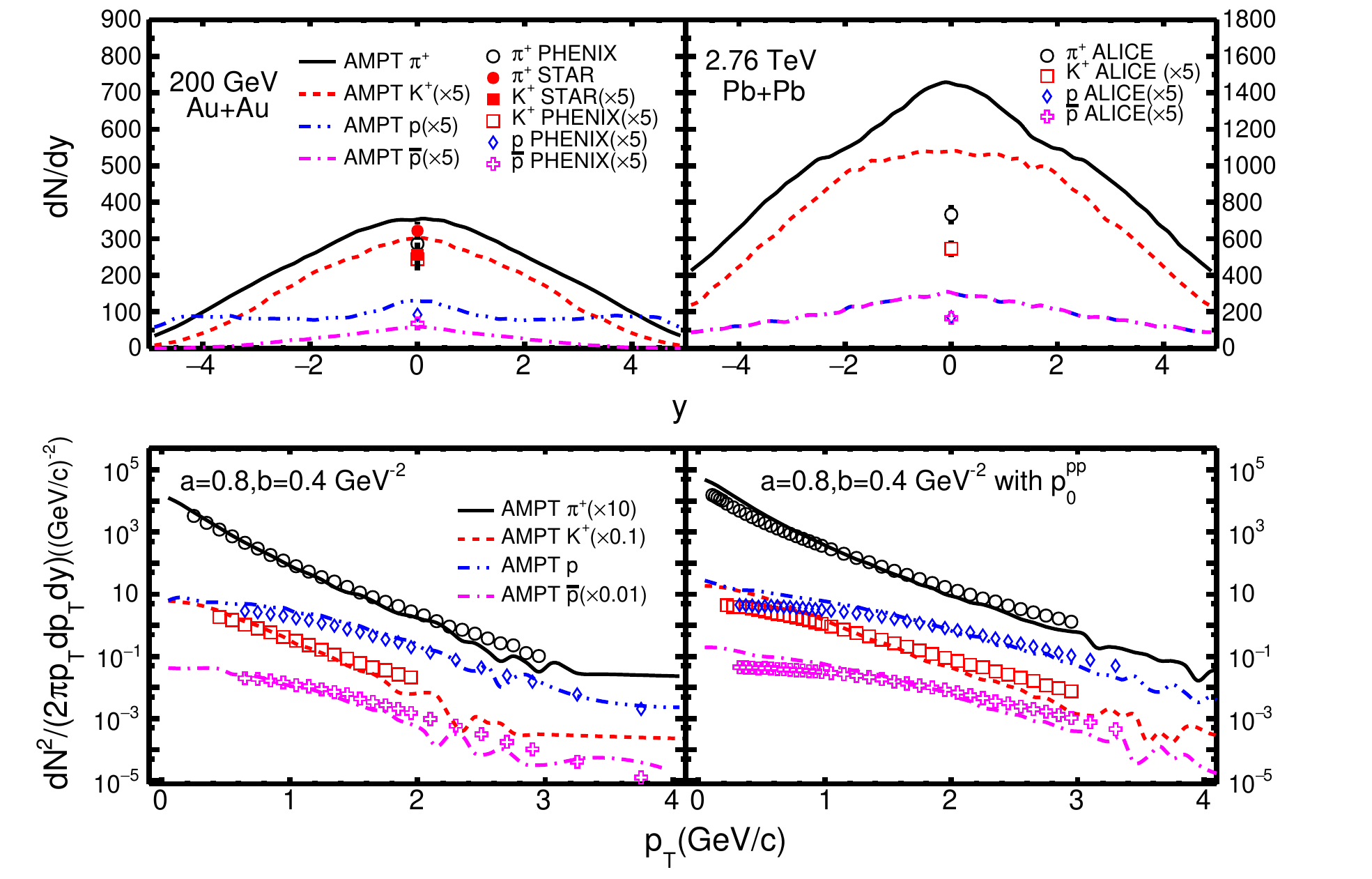}
   \caption{Identified particle $dN/dy$ distributions (upper panels)
 and $p_{\rm T}$ spectra (lower panels) for 0-5\% central Au+Au
 collisions at 200 GeV (left panels) and 0-5\% central Pb+Pb
 collisions at 2.76 TeV (right panels). 
Curves represent the AMPT results 
    using the same Lund fragmentation parameters 
and $p_0$ as for $pp$ collisions, while symbols represent experimental
data~\cite{Adler:2003cb,Abelev:2013vea}.
}
   \label{fig:pikpAA_untune}
  \end{center}
 \end{figure*}

Now we investigate results from the updated AMPT model 
on particles productions in nucleus-nucleus collisions. 
First we take the same parameters as for $pp$
collisions, i.e., Lund fragmentation parameters $a=0.8$,
$b=0.4$~GeV$^{-2}$, and the $p_0^{pp}(s)$ minijet cutoff function. 
Figure~\ref{fig:pikpAA_untune} shows the $dN/dy$ (left panels) and
$p_{\rm T}$ spectra (right panels) of $\pi^{+}, K^{+}, p$ and $\bar{p}$ 
for $0-5\%$ central Au+Au collisions at $\sqrt {s_{\rm NN}}=200$~GeV 
and $0-5\%$ central Pb+Pb collisions at 2.76~TeV, 
where results from the updated AMPT model 
are being compared to the experimental data
~\cite{Adler:2003cb,Abelev:2008ab,Abelev:2013vea}.  
Note that we show the PHENIX proton and antiproton data because they  
have been corrected for feed-down effects.
Also, the kaon and (anti)proton $dN/dy$ values 
from both the model and the experimental data have 
been multiplied by a constant factor for easier identification.

We see from Fig.~\ref{fig:pikpAA_untune} that 
the updated AMPT model with $a=0.8$, $b=0.4$~GeV$^{-2}$ 
and the $p_0^{pp}(s)$ minijet cutoff significantly overestimates the
yields of most of these particles for central heavy ion collisions at
both RHIC and LHC energies. 
Also, the $p_{\rm T}$ spectra of these particles from the AMPT model are
mostly softer than the data for both collision systems.
Moreover, with the $p_0^{pp}$ minijet cutoff and EPS09sNLO nuclear
shadowing, we find it impossible to reproduce the overall particle
yields of Pb+Pb collisions at LHC energies regardless of the Lund $a$
and $b$ values.

\begin{figure*}[!htb]
 \begin{center}
   \includegraphics[scale=0.9]{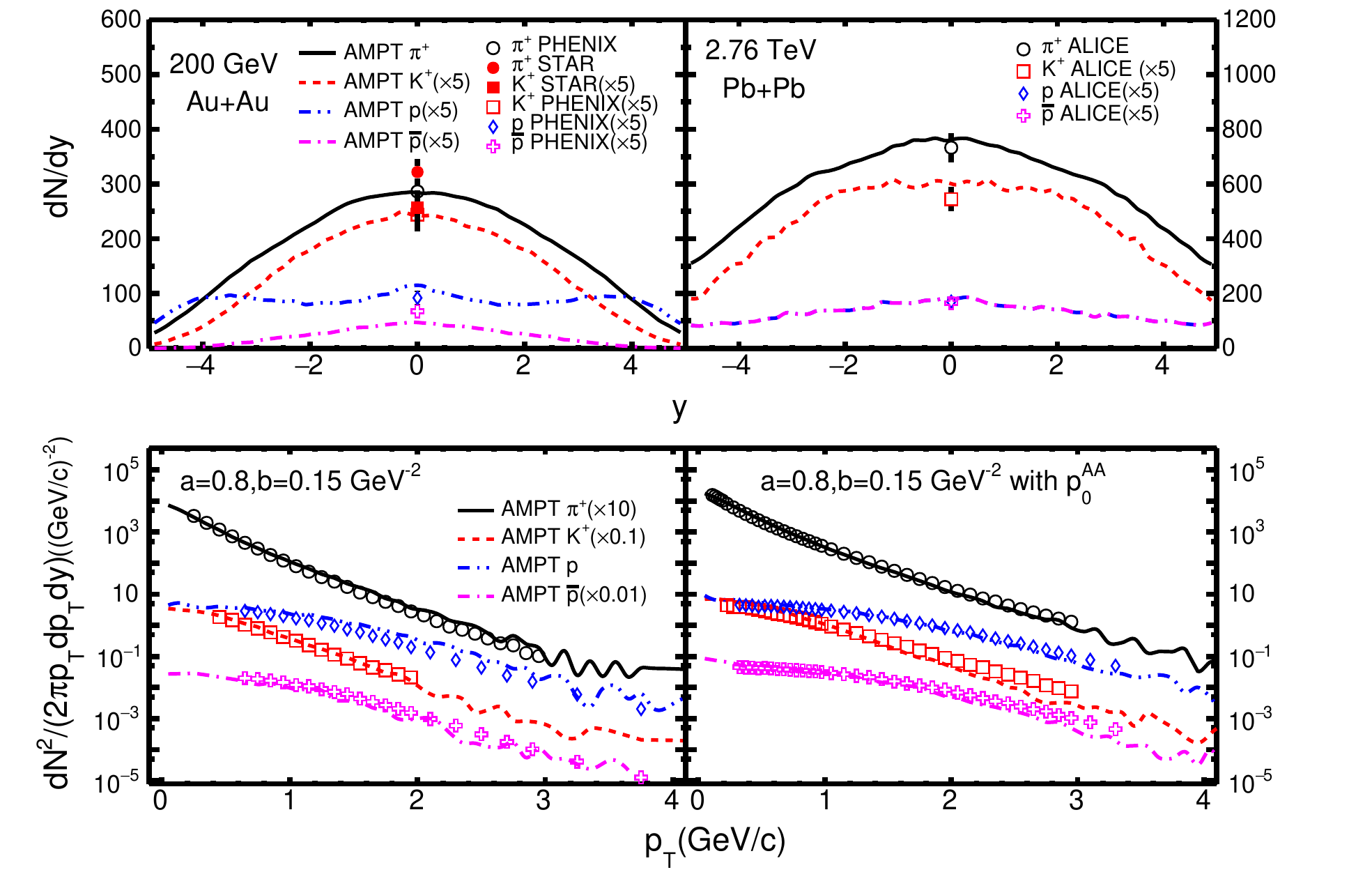}
   \caption{Same as Fig.~\ref{fig:pikpAA_untune} but using Lund
     $b=0.15$~GeV$^{-2}$ and $A$-scaled minijet cutoff $p_0^{AA}(s)$ for
     central $AA$ collisions; note however that $p_0^{AA}=p_0^{pp}$ at
     $\sqrt {s_{\rm NN}}=200$~GeV.
}
   \label{fig:pikp_tune}
  \end{center}
 \end{figure*}

We thus introduce the following $A$-scaling of
$p_0$, which increases the minijet cutoff $p_0$ for central AA
collisions at high energies such as the LHC:
\begin{eqnarray}
\label{eq:p0aa}
p_0^{AA}=&&p_0^{pp} A^{q(s)}, \nonumber \\ 
 q(s)=&&0.0334~ln \left ({\sqrt{s} \over 200} \right )-0.00232~ln^2
        \left ({\sqrt{s} \over 200} \right ) \nonumber \\ 
+&&0.0000541~ln^3 \left ({\sqrt{s} \over 200} \right ),
{\rm for~}\sqrt {s} \ge 200{\rm~GeV.}
\end{eqnarray}
In the above, $\sqrt{s}$ refers to $\sqrt{s_{\rm NN}}$ in $AA$
collisions and is in the unit of GeV. 
This $q(s)$ fit function is shown in Fig.~\ref{fig:p0_sigma}, 
where it is zero at $\sqrt {s_{\rm NN}} \le 200$ GeV, reaches a
value of 0.13 at $\sqrt {s_{\rm NN}}=10^5$ GeV, 
and approaches 0.16 at $\sqrt {s_{\rm NN}}\sim 10^7$ GeV.
The above nuclear scaling of the minijet momentum cutoff scale $p_0$ is 
motivated by the physics of color glass
condensate~\cite{McLerran:1993ni}, where the saturation momentum scale
$Q_s$ depends on the nuclear size as $Q_s \propto A^{1/6}$ in the
saturation regime for small-$x$ gluons in $AA$ collisions at
high-enough energies. 

We have decided to keep using the EPS09s nuclear shadowing, 
although it has significant uncertainties on its gluon shadowing
function at small $x$~\cite{Eskola:2009uj}. 
We also use the same Lund $a$ value of 0.8 for $AA$ collisions as for
$pp$ collisions, unlike in studies with the previous AMPT
model~\cite{Lin:2004en,He:2017tla}.
In addition, we find that a significantly smaller value for the Lund
$b$ parameter, $b=0.15$~GeV$^{-2}$, is needed to describe particle 
productions in $AA$ collisions.
This was also the case for the previous string melting version of the
AMPT model ~\cite{Lin:2014tya,Ma:2016fve}. 
Note that  throughout this study we use the default PYTHIA value of
0.30 for the relative production of strange to nonstrange quarks,
instead of imposing an upper limit of 0.40 as done for the 
string melting version of the previous AMPT model~\cite{Lin:2014tya}. 

Figure~\ref{fig:pikp_tune} shows the $dN/dy$ distributions  (left panels) 
and $p_{\rm T}$ spectra (right panels) from the AMPT model using
the new Lund $b$ parameter and $p_0^{AA}(s)$ cutoff in comparison with
the experimental data. We see that most of the $dN/dy$
data of $\pi^{+}, K^{+}, p$ and $\bar{p}$ in these central heavy ion
collisions can now be reasonably reproduced. 
The $p_{\rm T}$ spectra are also much harder than those in
Fig.~\ref{fig:pikpAA_untune} and mostly consistent with the
corresponding heavy ion data, due to the smaller value of the Lund $b$
parameter~\cite{Lin:2014tya}.  

\begin{figure}[!htb]
 \begin{center}
  \includegraphics[scale=0.43]{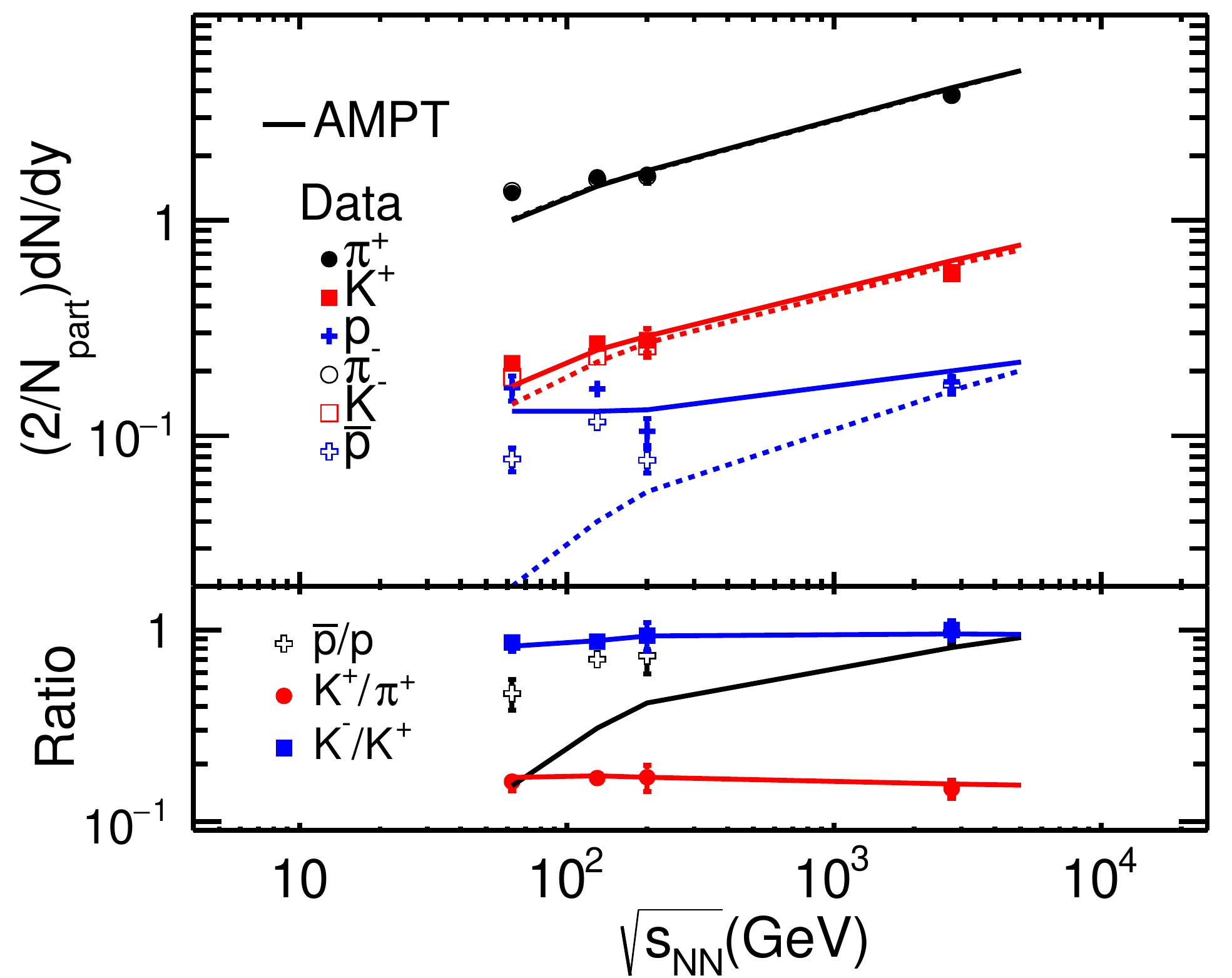}
   \caption{Identified particle $dN/dy$ 
and particle ratios at mid-rapidity in 0-5\% central $AA$ collisions 
versus the colliding energy; curves represent AMPT results, 
while symbols represent experimental data at 62.4
GeV~\cite{Abelev:2008ab}, 130 GeV~\cite{Adcox:2001mf}, 200
GeV~\cite{Adler:2003cb} and 2.76 TeV~\cite{Abelev:2013vea}.
}
   \label{fig:pikp_AA}
  \end{center}
 \end{figure}

In Fig.~\ref{fig:pikp_AA}, the energy dependences of identified
particle yields at mid-rapidity are shown in the upper panel for 0-5\%
central Au+Au collisions at RHIC energies and 0-5\% central Pb+Pb
collisions at LHC energies. 
The corresponding particle ratios are shown in the lower panel.  
Note that the rapidity range at 2.76~TeV is $|y|< 0.5$ 
while at other energies is $|y|< 0.1$,  
and that the PHENIX (anti)protons data at 62.4 and 130~GeV are not
corrected for feed-down from weak decays.
We see from Fig.~\ref{fig:pikp_AA} that the yields of charged pions
and kaons as well as their ratios  are well reproduced by the
updated AMPT model. 
However, similar to the trend in $pp$ collisions, 
at lower energies the string melting AMPT model underestimates the
anti-proton yield but tends to 
overestimates the proton yield at mid-rapidity. 
As a result, the mid-rapidity $\bar p/p$ ratios from the string
melting AMPT model at the lower RHIC energies are significantly
smaller than the experimental data. On the other hand, 
the AMPT model can reasonably reproduce the (anti)proton data for
central Pb+Pb collisions at the LHC energy of 2.76~TeV. 
These features are similar to those in the earlier study that used the
previous string melting AMPT model with the new quark coalescence  
~\cite{He:2017tla}.

\section{Discussions}
\label{sec:discussions}

\begin{figure}[!htb]
 \begin{center}
   \includegraphics[scale=0.43]{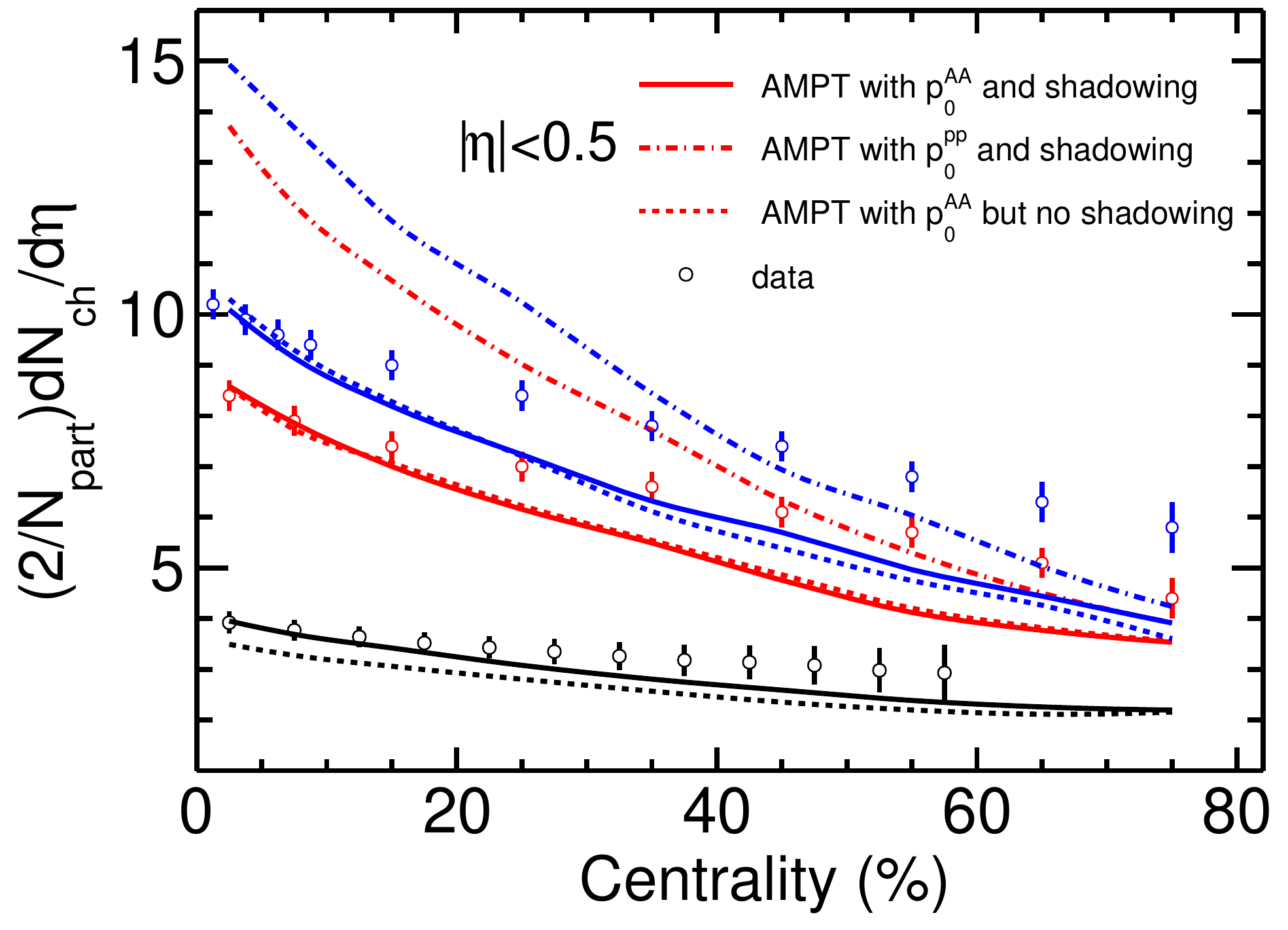}
   \caption{ Centrality dependence of charged particle  $dN/d\eta$
at mid-rapidity divided by $N_{\rm part}/2$ from the AMPT model 
for Au+Au collisions at 200~GeV and Pb+Pb collisions at 2.76~TeV and 
5.02~TeV with (solid) and without (dashed) nuclear shadowing in
comparison with the experimental data 
\cite{Aamodt:2010cz,Adare:2015bua,Adam:2015kda,Adam:2016ddh}. 
Dot-dashed curves represent AMPT results without using $A$-scaling
of $p_0$ that are more applicable to peripheral $AA$ collisions.
   }
   \label{fig:centrality_dep}
  \end{center}
 \end{figure}

Since the EPS09s nuclear shadowing is impact parameter-dependent and
diminishes for nucleons near the edge of the nucleus, we expect the
effect of nuclear shadowing to depend on centrality and
vanish for very peripheral $AA$ collisions. 
This is shown in Fig.~\ref{fig:centrality_dep} by the 
centrality dependence of charged particle $dN/d\eta$ 
within $|\eta| < 0.5$ divided by $N_{\rm part}/2$ 
for Au+Au collisions at 200~GeV and Pb+Pb collisions at 2.76~TeV 
and 5.02~TeV. 
Note that centrality is determined according to the
number of charged particles detected by the Beam-Beam Counters that
cover $3.0<|\eta|<3.9$ at 200~GeV 
or by the V0 detectors that cover $2.8<\eta<5.1$ and
$-3.7<\eta<1.7$ at 2.76~TeV or 5.02~TeV. 
The same centrality criterion is used in the analysis of our model
results, and we take $N_{\rm part}$ as the total number of nucleon
participants from both the projectile and target nuclei due to
inelastic collisions in the AMPT calculations. 

As expected, we see in Fig.~\ref{fig:centrality_dep} that 
the shadowing effect is very small for peripheral collisions. 
Actually, the figure shows that nuclear shadowing has a small effect
on charged particle yields at all centralities for $AA$ collisions
from the top RHIC energy to LHC energies. 
This is because of the large $p_0$ value at high energies and the weak
EPS09sNLO nuclear shadowing at large $Q^2$ values (that are at least
${p_0^{AA}}^2$) as shown in Fig.~\ref{fig:shadowing}. 

On the other hand, Fig.~\ref{fig:centrality_dep} shows that the 
$A$-scaling of $p_0$ has a large effect on charged particle yields in 
$AA$ collisions at LHC energies, especially for more central
collisions. 
As mentioned earlier, the string melting AMPT model significantly
overestimates the charged particle yields 
in central Pb+Pb collisions at LHC when it uses $p_0^{pp}$, the
same minijet cutoff scale as for $pp$ collisions. 
After the $A$-scaling of $p_0$, however, the minijet cutoff scale in $AA$
collisions ($p_0^{AA}$) at LHC energies becomes significantly higher
and thus $\sigma_{\rm jet}$ becomes much smaller, as shown in
Fig.~\ref{fig:parameterfit} by the dashed line that is much lower than
the dotted line at LHC energies. This leads to a significant
decrease of the charge particle yields at LHC energies, especially for
central $AA$ collisions where the binary scaling of minijet
productions makes them more sensitive to the minijet cutoff $p_0$.

For peripheral $AA$ collisions however, we expect no need for the
$A$-scaling of $p_0$, because participant nucleons there are near the
edge of the nucleus and should be almost free of saturation effects. 
Since we have not implemented this impact parameter-dependent nuclear
scaling of $p_0$ and the current $A$-scaling of Eq.~(\ref{eq:p0aa}) is
only valid for central $AA$ collisions, 
we show in Fig.~\ref{fig:centrality_dep} the LHC Pb+Pb results without
using the $A$-scaling of $p_0$ (dot-dashed lines), which are more
suitable for peripheral collisions. 
Indeed, we see that the AMPT results without the $A$-scaling of $p_0$
give higher charged particle yields and are closer to the experimental
data for peripheral collisions than the AMPT results with $A$-scaling
of $p_0$. 
Also note that, since we have found that the Lund $b$ value is much
smaller in central $AA$ collisions than in $pp$ collisions, 
the Lund $b$ value should depend on the system size or 
centrality, and increasing its value for peripheral $AA$ collisions
(similar to $pp$ collisions) could further improve the description of
charged particle yields there.

We have seen that the minijet cutoff scale $p_0$ becomes increasingly
large with energy and can be more than 4 or even 6 GeV$/c$. 
However, it is questionable to treat transverse momentum exchanges
below such a high value of $p_0$ as soft physics with the Lund
string fragmentation, while the production of charm particles is
usually viewed as a perturbative-QCD process where the FONNL approach
has been very successful~\cite{Cacciari:2012ny}.
Therefore the two-component model such as HIJING may be
problematic for the initial condition at very high energies. 
For example, the need for us to introduce the nuclear scaling of $p_0$
for $AA$ collisions at LHC energies and above may indicate the  
importance of saturation physics for large systems at very high
energies. In addition, the current parton cascade in the AMPT
model only includes elastic parton scatterings~\cite{Zhang:1997ej}. 
However, inelastic parton interactions~\cite{Xu:2004mz} affect the
parton abundance and momentum spectrum at high energies, and these
effects are expected to be energy- and centrality-dependent. 
Therefore including inelastic parton scatterings should improve the
physics of a multi-phase transport model~\cite{Lin:2014uwa}.

The updated AMPT model has not shown obvious phenomenological
improvements over the previous AMPT model when compared with the
experimental data in this study, 
except that the updated model uses the same Lund $a$ value
for $pp$ and $AA$ collisions at all energies and thus removes the 
uncertainty of this parameter present in the previous AMPT model. 
However, the updated AMPT model should be more robust in its physics
because of its inclusion of modern parton PDFs in the nuclei. 
Therefore we expect it to provide a better foundation for future model 
developments and also show improvements in certain observables such as 
heavy flavor productions~\cite{charm}. 

\section{summary}
\label{sec:summary}

A multi-phase transport model has been using the old Duke-Owens parton
distribution functions for the free proton and a schematic nuclear 
shadowing parameterization.  
This leads to significant uncertainties in its ability to
address heavy flavor and/or high-$p_{\rm T}$ particles, because they are
produced by perturbative-QCD processes and thus directly depend on the parton
distribution functions of nuclei. 
In this study, we have incorporated a modern set of free proton parton
distribution functions, the CTEQ6.1M set, and the impact
parameter-dependent EPS09sNLO nuclear shadowing in an updated AMPT
model.
We first determine the energy dependence of two key parameter
functions, $p_{0}(s)$ and $\sigs(s)$, in the HIJING two-component
model by fitting the experimental data on total and inelastic cross
sections of $pp$ and $p \bar p$ collisions from $\sqrt s \sim $ 4 GeV
to 13 TeV.  
We then compare particle productions from the string melting
version of the updated AMPT model with the experimental data in 
both $pp$ and $AA$ collisions at RHIC and LHC energies. 
We find that the $p_{0}(s)$ function and the constant values for the
Lund string fragmentation parameters that can reasonably describe the
particle yields and $p_{\rm T}$ spectra in $pp$ collisions fail to
describe central $AA$ collisions at LHC energies.  
Therefore we introduce a nuclear scaling of the minijet transverse
momentum cutoff $p_0$ for central $AA$ collisions at high energies
that is motivated by the color glass condensate picture. 
Then the string melting AMPT model can also reasonably describe the
overall particle yields and $p_{\rm T}$ spectra of $AA$ collisions at
both RHIC and LHC energies. 
We expect the updated AMPT model to provide more reliable
descriptions of heavy flavor and high-$p_{\rm T}$ observables in
relativistic collisions of both small and large systems. 
It also serves as a good foundation for further improvements of the model. 
 
\begin{acknowledgments}
This work is supported by the MoST of China 973-Project
No. 2015CB856901 (FL \& SS) and the National Natural Science
Foundation of China under grant No. 11890711  and 11628508 (ZWL, FL
\& SS).
\end{acknowledgments}

\end{document}